\documentclass[sigconf,nonacm]{acmart}

\usepackage{multirow}
\usepackage{booktabs}
\usepackage[table]{xcolor}
\usepackage{algorithm}
\usepackage{algpseudocode}
\usepackage{enumitem} 
\usepackage{booktabs,multirow,bm}
\usepackage{tcolorbox}
\usepackage{listings}
\usepackage[normalem]{ulem}
\AtBeginDocument{
  }

\setcopyright{none}
\acmDOI{}
\acmISBN{}

\begin{document}

\title{Visual Graph Reasoning via Knowledge Compilation}
\author{Rongzheng Wang}
\email{wangrongzheng@std.uestc.edu.cn}
\affiliation{
  \institution{University of Electronic Science and Technology of China}
  \city{Chengdu}
  \country{China}}

\author{Zhe Wang}
\email{wzhe@std.uestc.edu.cn}
\affiliation{
  \institution{University of Electronic Science and Technology of China}
  \city{Chengdu}
  \country{China}}

\author{Ke Qin}
\authornotemark[2]
\email{qinke@uestc.edu.cn}
\affiliation{
  \institution{University of Electronic Science and Technology of China}
  \city{Chengdu}
  \country{China}}

\author{Rongwei Wang}
\email{wrw24@mails.tsinghua.edu.cn}
\affiliation{
  \institution{Tsinghua Shenzhen International Graduate School}
  \city{Shenzhen}
  \country{China}}

\author{Muquan Li}
\email{muquanli831@foxmail.com}
\affiliation{
  \institution{University of Electronic Science and Technology of China}
  \city{Chengdu}
  \country{China}}

\author{Yizhuo Ma}
\email{myz@std.uestc.edu.cn}
\affiliation{
  \institution{University of Electronic Science and Technology of China}
  \city{Chengdu}
  \country{China}}

\author{Yihong Huang}
\email{2021080910004@std.uestc.edu.cn}
\affiliation{
  \institution{University of Electronic Science and Technology of China}
  \city{Chengdu}
  \country{China}}

\author{Jielei Wang}
\authornotemark[2]
\email{jieleiwang\_uestc@163.com}
\affiliation{
  \institution{University of Electronic Science and Technology of China}
  \city{Chengdu}
  \country{China}}

\author{Shuang Liang}
\authornote{Corresponding author.}
\additionalaffiliation{
  \institution{Ubiquitous Intelligence and Trusted Services Key Laboratory of Sichuan Province}
  \city{Chengdu}
  \country{China}}
\email{shuangliang@uestc.edu.cn}
\affiliation{
  \institution{University of Electronic Science and Technology of China}
  \city{Chengdu}
  \country{China}}

\renewcommand{\shortauthors}{Rongzheng Wang et al.}

\begin{abstract}
Visual graph reasoning requires answering graph-theoretic questions directly from graph images, where graph topology and state are conveyed visually rather than given in symbolic form. Despite recent progress of vision-language models (VLMs), current approaches to visual graph reasoning still fail on simple visual graph problems. This reveals a fundamental limitation of existing approaches: they prioritize final-answer supervision over the intermediate recovery of an explicit graph representation that preserves graph topology and state from visual input. To address this limitation, we propose \textbf{VGCompiler}, a compilation-centric paradigm for visual graph reasoning via knowledge compilation. VGCompiler organizes reasoning around two compilers: a representation compiler that recovers a structure-preserving intermediate graph representation from visual input, and an operation compiler that compiles query intent under the recovered graph state into an executable graph operation. Specifically, we build VGCompiler on Qwen3-VL-8B and train it with reinforcement learning guided by a layered reward over executability, compiled graph validity, representation quality, and operation quality. VGCompiler uses a frozen observer to summarize graph and question conditions into lightweight signatures, enabling archive retrieval and code reuse across similar regimes. Experiments on three benchmarks GVLQA, VisionGraph, and VGCURE, show that Qwen-VGCompiler, built on an 8B backbone, surpasses the strongest closed-source VLM baseline by 28.9\% and the strongest code-based baseline by 23.7\%, while maintaining high efficiency. We further evaluate VGCompiler on three real-world domains, including metro routing, logistics delivery, and network fault assessment, where it generalizes across heterogeneous visual graphs and domain-grounded tasks.
\end{abstract}



\maketitle

\section{Introduction}

\begin{figure}[t]
\centering
\setlength{\abovecaptionskip}{2pt}
\setlength{\belowcaptionskip}{0pt} 
\includegraphics[width=0.95\linewidth]{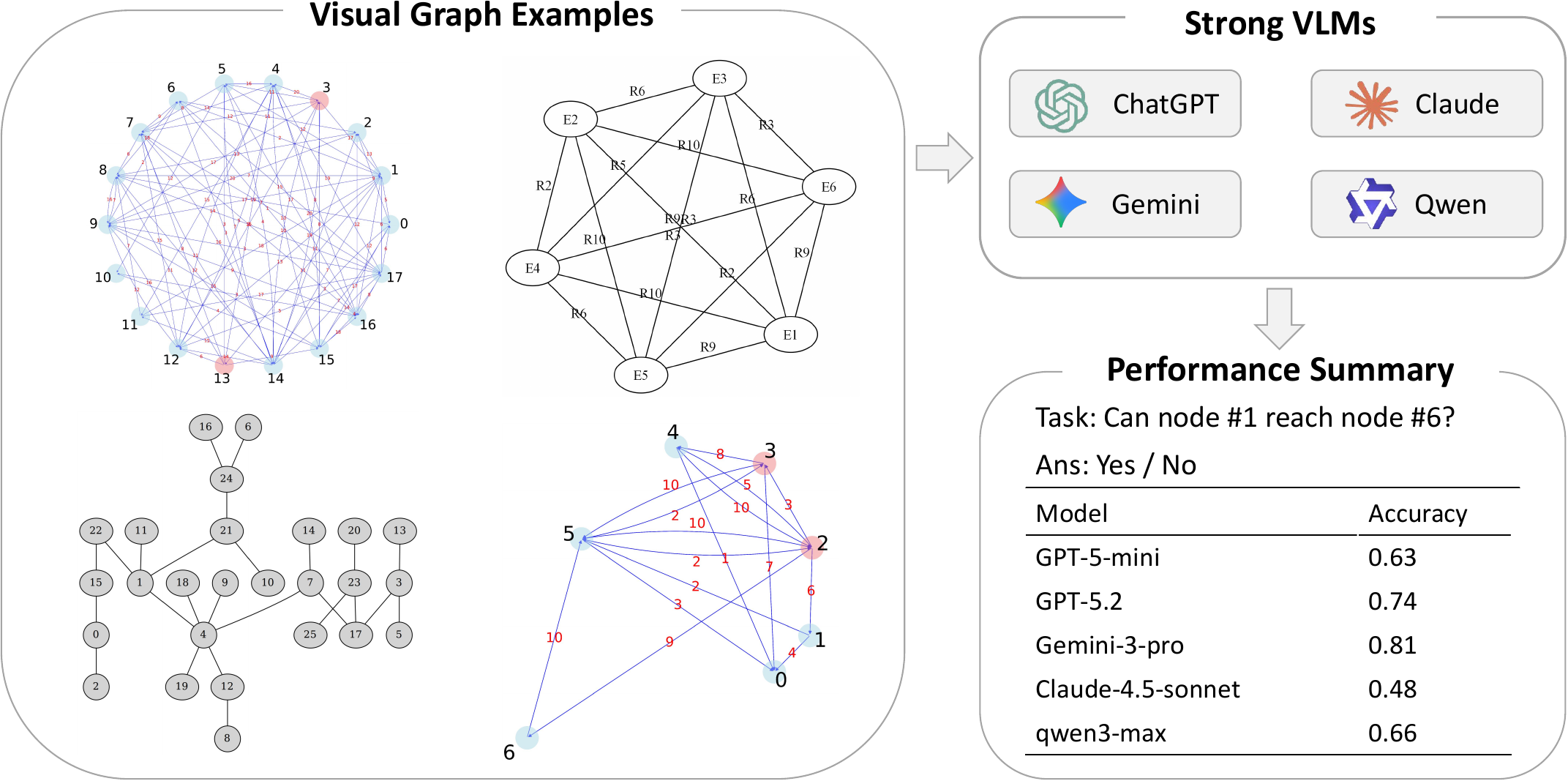}
\caption{Representative visual graph examples together with a simple reachability evaluation across mainstream VLMs.}
\label{fig:intro_graph_table}
\vspace{-0.5em}
\end{figure}

Vision-language models (VLMs)~\cite{clip2021, flamingo2022, blip22023,instructblip2023, llava15} have shown strong capability on broad multimodal reasoning benchmarks~\cite{mmmu2024,mathvista2024}, yet current VLM-based methods still struggle with even simple visual graph reasoning tasks~\cite{gita2024,visiongraph2024,vgcure2025}. Visual graph reasoning asks a model to answer graph queries such as reachability, cycle existence, and shortest path directly from graph images, where graph topology and state are conveyed visually rather than given in symbolic form~\cite{gita2024,visiongraph2024}. Reliable solving therefore depends on recovering an explicit graph representation from visual input. However, existing approaches often prioritize the final answer over this intermediate recovery, so they may still produce plausible but incorrect answers from partial or incorrect graph representations. Figure~\ref{fig:intro_graph_table} illustrates this failure: even for a simple reachability query, strong VLMs remain unreliable across visually diverse graph renderings.

This failure reveals two explicit limitations. (1) \textit{Graph style affects representation recovery}. Unlike text graph reasoning, where relational structure is directly available in symbolic or textual form, visual graph reasoning must recover a faithful graph representation from visually varied renderings. Benchmarks such as Visual Graph Arena~\cite{visualgrapharena2025} show that severe errors can persist even when graph semantics remain unchanged and only layout or style is altered, indicating that visually irrelevant factors can destabilize this recovery. (2) \textit{Graph state affects downstream operation}. Even with a recovered graph representation, solving a graph query still requires an operation consistent with the graph state. For example, shortest-path reasoning~\cite{bellman1958routing,dijkstra1959connexion,floyd1962shortest} over unweighted graphs, nonnegative-weighted graphs, and graphs with negative edges relies on different executable procedures~\cite{warshall1962boolean,tarjan1972dfs}. Text-based methods such as GraphTool-Instruction~\cite{graphtoolinstruction2025} and GraphCogent~\cite{graphcogent2025} assume explicit graph states and therefore select downstream operations more directly, whereas visual graph reasoning must infer graph state first. Consequently, early state-recognition errors directly propagate into downstream execution.

These two limitations originate from the same mismatch: visual graphs present graph semantics in a visual form whose appearance varies with style, layout, and rendering, whereas graph queries require an explicit graph representation together with a valid downstream operation. Visual graph reasoning therefore requires recovering a faithful graph representation from the input and identifying an operation consistent with the recovered graph state. This makes it a knowledge compilation problem~\cite{darwiche2002kcmap}, where the visual input acts as a weak source language and is mapped into a target representation that preserves graph semantics and supports downstream operations. This view reframes visual graph reasoning into two compilers: a representation compiler that compiles the visual graph into a structure-preserving graph representation, and an operation compiler that determines the executable graph operation required by the query under the recovered graph state.

Inspired by the knowledge compilation view, we propose \textbf{VGCompiler}, a compilation-centric paradigm for visual graph reasoning via knowledge compilation. In VGCompiler, executable code serves as the basic unit of learning and reuse, and organizes reasoning around two compilers: the representation compiler maps a visual graph into a structure-preserving intermediate representation; the operation compiler determines the executable graph operation required under the recovered graph state. To learn code that adapts to different graph-question conditions, VGCompiler uses a VLM as the code generator and trains it with reinforcement learning to generate candidate code conditioned on the current graph-question pair. The optimization is driven by a layered reward over executability, compiled graph validity, representation quality, and operation quality. To archive similar graph-question pairs and enable code reuse, VGCompiler employs a frozen observer to extract lightweight cues from each pair. These cues serve as compact condition signatures for organizing archived code during training and retrieving the most compatible archived code during inference.

\begin{figure}[t]
\centering
\setlength{\abovecaptionskip}{2pt}
\setlength{\belowcaptionskip}{0pt} 
\includegraphics[width=0.99\linewidth]{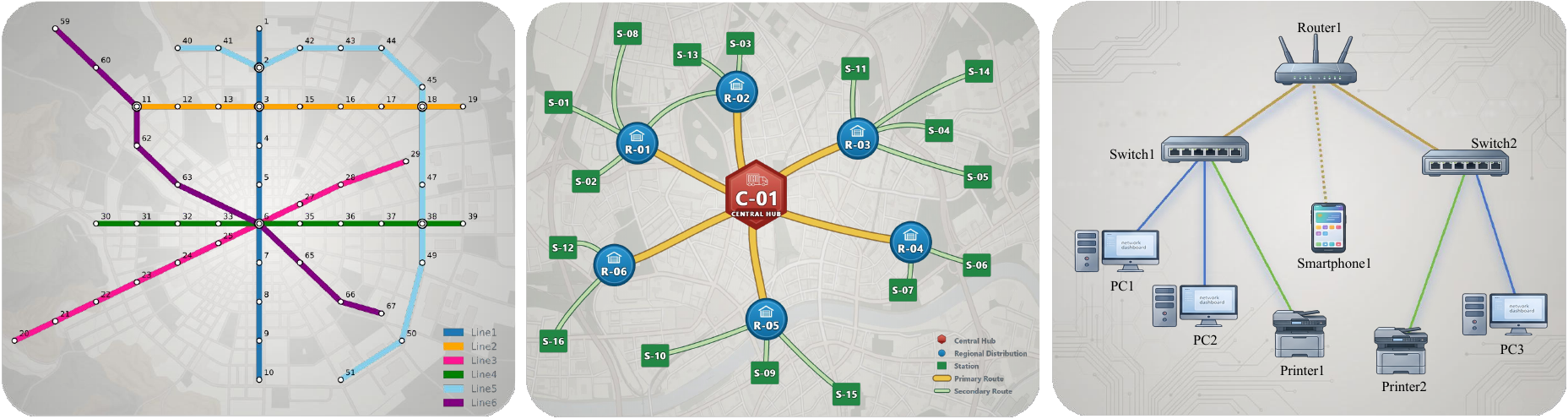}
\caption{Three representative real-world visual graph reasoning scenarios: metro routing, logistics delivery, and network fault assessment.}
\label{fig:real_world_reasoning_examples}
\vspace{-1em}
\end{figure}

We further fine-tune Qwen3-VL-8B~\cite{qwen3vl8b} with reinforcement learning to obtain Qwen-VGCompiler. Across three public benchmarks, GVLQA~\cite{gita2024}, VisionGraph~\cite{visiongraph2024}, and VGCURE~\cite{vgcure2025}, Qwen-VGCompiler demonstrates strong effectiveness and generalization, outperforming prior state-of-the-art methods by 28.9\%. We further evaluate VGCompiler on three real-world scenarios, as illustrated in Figure~\ref{fig:real_world_reasoning_examples}: metro routing, logistics delivery, and network fault assessment. Across these heterogeneous settings, VGCompiler also demonstrates real-world generalization.

Our contributions are summarized as follows: (1) We formulate visual graph reasoning under a knowledge compilation view, where solving a visual graph query requires both recovering a structure-preserving graph representation from visual input and identifying a valid downstream operation under the recovered graph state; (2) We propose VGCompiler, a compilation-centric paradigm that instantiates this view through VLM-generated executable code. VGCompiler comprises a representation compiler and an operation compiler, together with a reusable code archive indexed by lightweight graph-question signatures for retrieval and reuse; (3) We introduce a layered reward over executability, compiled graph validity, representation quality, and operation quality to guide reinforcement learning, and we validate strong effectiveness on public benchmarks together with real-world generalization across heterogeneous visual graph scenarios.

\section{Related Work}

\subsection{LLMs for Graph Reasoning}

LLM-based graph reasoning was first studied mainly in text-based settings, where the graph is already given in symbolic form and the main challenge lies in selecting and executing the correct graph operation. Representative methods strengthen this process through explicit programs~\cite{dash2026}, tool use~\cite{topollm2026}, and multi-step collaboration, including Chain-of-Thought prompting~\cite{wei2022cot}, Self-Consistency~\cite{wang2023selfconsistency}, PAL~\cite{gao2023pal}, ReAct~\cite{yao2023react}, and Toolformer~\cite{schick2023toolformer}. Attention-based mechanisms offer another direction for graph reasoning by improving long-context information selection and adaptively controlling reasoning generation~\cite{dsas2025,asag2026}. Graph-specific systems such as GraphWiz~\cite{graphwiz2024}, GraphTool-Instruction~\cite{graphtoolinstruction2025}, GraphCogent~\cite{graphcogent2025}, and NeuroPath~\cite{neuropath2025} further adapt these ideas to graph reasoning. However, these methods assume that graph topology and state are already externalized. More recently, graph reasoning has been extended to visual graph inputs, where the graph must first be recovered from images. GITA~\cite{gita2024}, VisionGraph~\cite{visiongraph2024}, VGCURE~\cite{vgcure2025}, and GraphVis~\cite{graphvis2024} move in this direction through graph-aware modeling, staged reasoning, or task-specific visual adaptation. However, current visual methods still weakly model the interface between graph recovery and downstream graph operation, especially under style variation and heterogeneous graph-state conditions~\cite{visualgrapharena2025}.

\subsection{Visual Graph Reasoning Benchmarks}

Beyond broad multimodal benchmarks such as MMMU~\cite{mmmu2024} and MathVista~\cite{mathvista2024}, recent work has begun to study visual graph reasoning as a distinct problem. Visual graph reasoning studies how to answer graph-theoretic questions when the graph is presented in visual form rather than as a symbolic input, where graph-structured information is widely conveyed through diagrams, maps, interfaces, and other visual media. GITA~\cite{gita2024} introduces GVLQA for vision-language graph reasoning, VisionGraph~\cite{visiongraph2024} expands the setting to a broader range of graph-theoretic tasks in visual contexts, and VGCURE~\cite{vgcure2025} further distinguishes graph understanding from graph reasoning. More recently, Visual Graph Arena~\cite{visualgrapharena2025} shifts the focus to robustness under layout and style variation, showing that visually irrelevant rendering changes can still alter model behavior even when graph semantics are preserved. These benchmarks indicate that topology recovery, graph-state grounding, and robustness to visual variation remain core challenges in visual graph reasoning.

\subsection{Knowledge Compilation}

Knowledge compilation studies how to map a source representation into a target language that preserves semantics while better supporting downstream queries and operations~\cite{darwiche2001dnnf,darwiche2002dnnfcompiler,darwiche2002kcmap}. Classical work focuses on symbolic forms such as BDDs~\cite{bryant1986bdd} and DNNF~\cite{darwiche1999dnnf}. This perspective is directly relevant to visual graph reasoning, where graph topology and state are visually encoded rather than explicitly given in symbolic form. We follow this view by compiling a visual graph into an attributed adjacency list that preserves graph semantics and supports downstream graph operations. From a complementary compression perspective, recent dataset-distillation methods retain task-relevant information through adaptive quantization, topology-aware dynamic retrieval, and efficient inner-loop optimization~\cite{DBLP:conf/aaai/LiZDXQ25,li2026fixed,DBLP:conf/nips/LiGZLXOQ25}, providing potential principles for scaling visual graph compilation while preserving the exact graph semantics required for downstream execution.

\begin{figure}[t]
\centering
\setlength{\abovecaptionskip}{2pt}
\setlength{\belowcaptionskip}{0pt} 
\includegraphics[width=0.9\columnwidth]{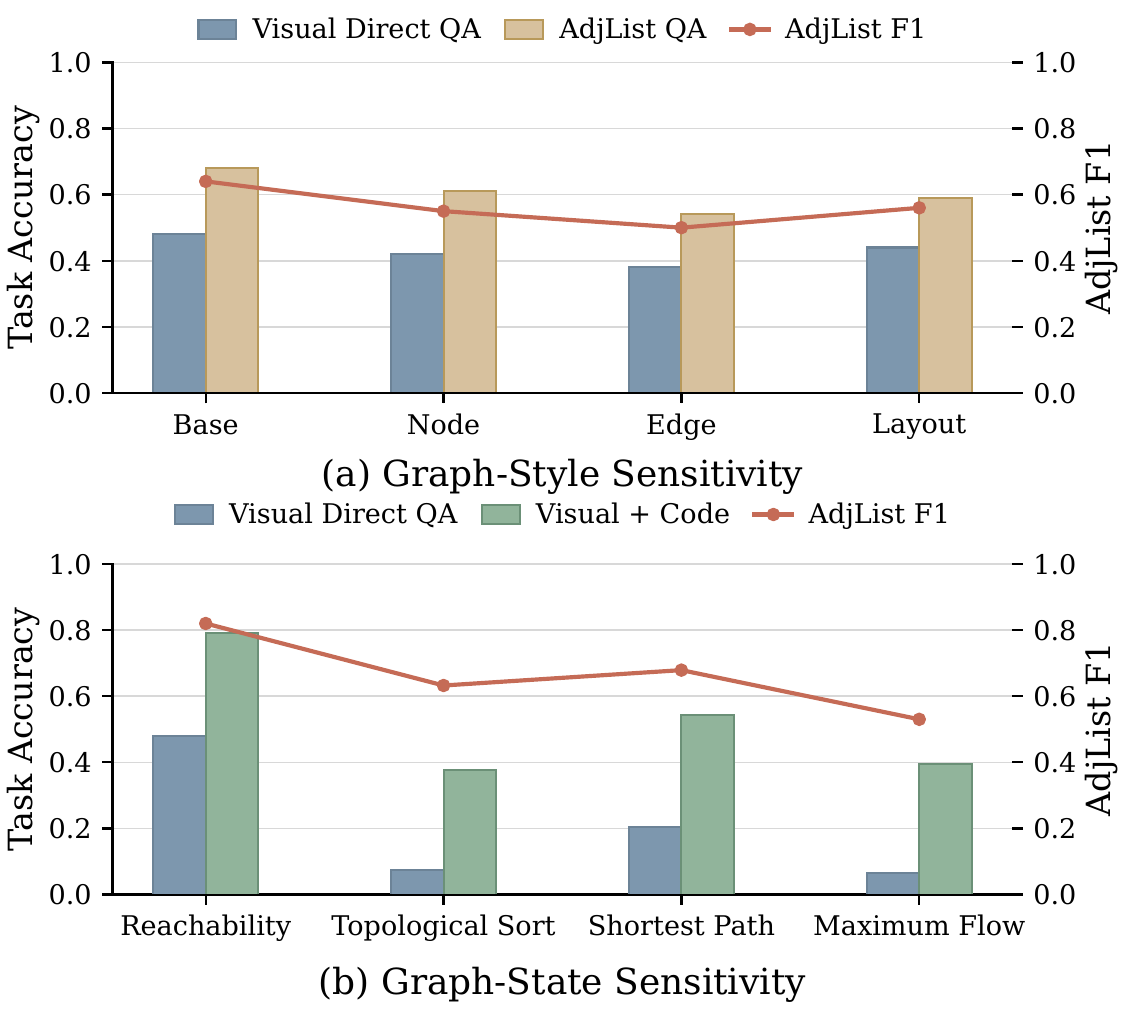}
\caption{Graph style and graph state sensitivity: (a) varies visual style while fixing the graph; (b) varies graph-state conditions and downstream operations while fixing style and layout. The results show that robust visual graph reasoning depends on both graph compilation and graph operation.}
\label{fig:graph_method}
\vspace{-0.5em}
\end{figure}

\section{Knowledge Compilation in Visual Graph Reasoning}

\subsection{Visual Graphs as a Weak Source Language}

The central challenge of visual graph reasoning lies in a mismatch between how graph semantics are presented and how graph queries should be solved. Visual graphs present graph semantics in a style-variable visual form, whereas graph queries require a faithful graph representation together with a valid downstream operation. Visual graph reasoning therefore requires recovering a reliable graph representation from the rendered input and identifying the valid downstream operation under the recovered graph state.

Knowledge compilation frames this mismatch as a gap between the visual source language in which the graph is presented and the target representation required for reliable downstream reasoning. Let $g(x)$ denote the graph semantics of image $x$, consisting of topology $t(x)$ and graph state $s(x)$. A target representation $z$ should faithfully encode these semantics and support the valid operations required by question $q$ under graph state $s(x)$. Let $\mathrm{Sem}(z)$ denote the graph semantics represented by $z$, and let $\mathcal O(q,s(x))$ and $\mathcal Q(z)$ denote the required and supported operation sets. We then require:
\begin{equation}
\mathrm{Sem}(z)=g(x),\qquad \mathcal O(q,s(x))\subseteq \mathcal Q(z),
\end{equation}
where $\mathcal O(q,s(x))$ is the set of valid operations induced by question $q$ under graph state $s(x)$, and $\mathcal Q(z)$ is the operation set supported by $z$. Visual graphs therefore form a weak source language for reasoning, since the underlying graph is not directly exposed in a form suitable for reliable downstream operation.

Figure~\ref{fig:graph_method}(a) studies graph style sensitivity by fixing the underlying graph topology and varying node style, edge style, and layout over 100 reachability queries from GVLQA. We compare three quantities: direct visual QA by Qwen3-VL-8B, QA with the complete adjacency list provided, and the F1 of the recovered adjacency list. Direct visual QA is sensitive to these perturbations, whereas QA with the explicit adjacency list remains more stable. The recovered adjacency F1 also drops under the same perturbations, indicating that style variation mainly degrades representation recovery and that the resulting errors then propagate to downstream reasoning.

These observations motivate a code-based formulation. In VGCompiler, the target representation is instantiated as an attributed adjacency list, which preserves graph semantics while exposing local structure and state attributes in executable form. We therefore use the attributed adjacency list as the target representation:
\begin{equation}
z=\left\{\left(v_i,\mathcal N(v_i)\right)\right\}_{i=1}^{n},\qquad
\mathcal N(v_i)=\left\{(v_j,a_{ij})\right\},
\end{equation}
with $n$ the number of nodes, $\mathcal N(v_i)$ the attributed neighbor set of $v_i$, and $a_{ij}$ encoding edge attributes such as direction and weight. This representation preserves graph semantics and provides a unified interface for downstream reasoning. Under this interface, visual graph reasoning can be organized around two compilers:
\begin{equation}
z=\mathrm{Comp}(x;c^{\mathrm{rep}}),\qquad
\hat y=\mathrm{Exec}(z,q;c^{\mathrm{op}}),
\end{equation}
where $c^{\mathrm{rep}}$ serves as the representation compiler, which maps visual evidence into an attributed adjacency list encoding graph semantics, and $c^{\mathrm{op}}$ serves as the operation compiler, which determines the executable graph operation required under the recovered graph state; this operation is then carried out on the compiled representation to produce the predicted answer $\hat y$.

\subsection{Representation and Operation Conditions}

Figure~\ref{fig:graph_method}(b) studies graph state sensitivity by fixing the rendering style and layout and comparing four graph-question settings with different graph state conditions (e.g., direction and edge weights) and downstream operations. We compare direct visual QA by Qwen3-VL-8B, code-based QA from the same visual input, and the F1 of adjacency-list recovery via code generation. Even under the same style and layout, the variation in adjacency-list F1 shows that graph state conditions still affect representation recovery. Although code-based QA outperforms direct visual QA, its accuracy still varies substantially across settings, indicating that graph state affects not only representation recovery but also downstream operation.

This observation reveals the second source of difficulty. Unlike style variation, which mainly affects representation recovery, graph-state conditions also affect how the recovered graph should be operated on. In settings such as involving direction or weights, small recovery errors are more likely to alter both the recovered representation and the validity of the downstream operation. Reliable solving therefore depends on faithful representation recovery and state-consistent execution.

\section{Method}

\subsection{Overview of VGCompiler}

Motivated by the above observations, we propose \textbf{VGCompiler}. Unlike previous paradigms that directly map graph images to answers, VGCompiler treats executable code as the basic unit of learning and reuse for visual graph reasoning. Each generated code instance consists of two coupled code blocks, \(c=(c^{\mathrm{rep}},c^{\mathrm{op}})\), where \(c^{\mathrm{rep}}\) is the representation compiler that maps a visual graph into a standardized attributed adjacency list, and \(c^{\mathrm{op}}\) is the operation compiler that determines the executable graph operation required under the recovered graph state.

\begin{figure*}[t]
\centering
\setlength{\abovecaptionskip}{2pt}
\setlength{\belowcaptionskip}{0pt} 
\includegraphics[width=0.98\textwidth]{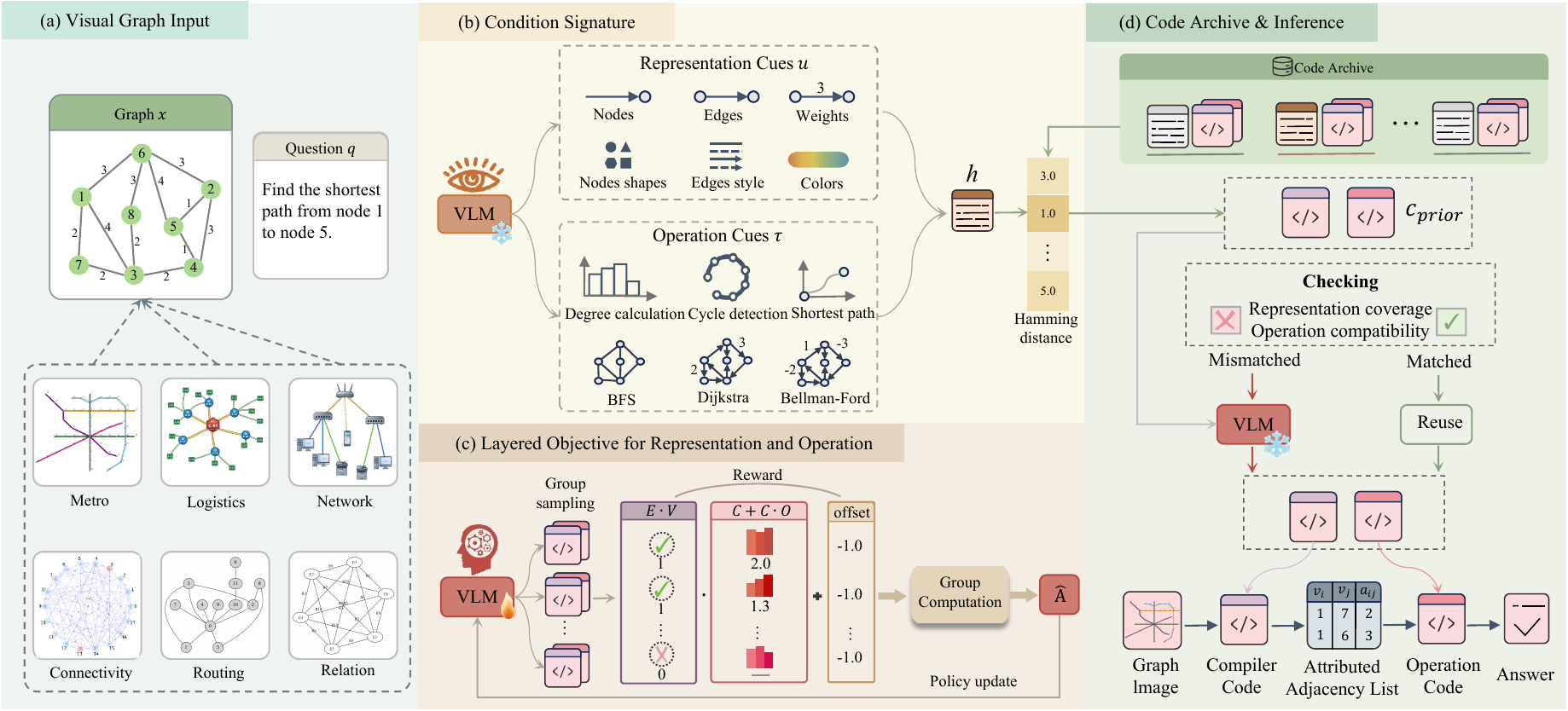}
\caption{Overview of VGCompiler. A frozen observer summarizes each graph-question pair into a lightweight condition signature. Guided by a layered reward, the code policy learns a paired code instance for representation recovery and downstream operation. At inference, the signature retrieves the nearest archived code prior and either reuses it directly or revises it into new code according to representation coverage and operation compatibility.}
\label{fig:vgcoder_main}
\end{figure*}

As shown in Figure~\ref{fig:vgcoder_main}, VGCompiler combines three core components: a frozen observer, a trainable code policy, and a code archive. The observer summarizes representation and operation cues from each graph-question pair into a lightweight condition signature; the code policy is trained with reinforcement learning guided by a layered reward over executability, compiled graph validity, representation quality, and operation quality; and the code archive accumulates reusable priors during training and provides retrievable priors for condition-aware code reuse at inference.

\subsection{Conditioned Code Generation}

A single code solution is not equally effective for all graph-question pairs. The representation compiler depends on visual graph style, whereas the operation compiler depends on both query intent and visible state cues such as direction or edge weights. We therefore condition code generation on the current graph-question pair. Given \((x,q)\), we freeze the VLM parameters and use it as an observer:
\begin{equation}
(u,\tau)=\mathrm{Obs}_{\psi}(x,q), \qquad
h=\mathrm{Sig}(u,\tau),
\end{equation}
where \(\psi\) denotes the frozen parameters of the observer VLM, \(u\) denotes \emph{representation cues}, i.e., lightweight textual descriptions of graph attributes relevant to representation recovery, including nodes, edges, edge weights, and their visual realization such as shape, heterogeneity, and color distinction. The variable \(\tau\) denotes \emph{operation cues}, namely lightweight descriptions of the operation requirements implied by the query together with visible state cues relevant to downstream execution. The signature \(h\) serves as a compact condition descriptor.

During inference, using Hamming distance over condition signatures, we retrieve the nearest archived code prior from the code archive \(\mathcal C^\star\):
\begin{equation}
s^\star=
\arg\min_{(h_s,\hat c_s,\bar R_s)\in\mathcal C^\star} d_H(h,h_s),
\qquad
c_{\mathrm{prior}}=\hat c_{s^\star},
\end{equation}
where \(s^\star\) is the retrieved archive index, and \(h_{s^\star}\) is the signature associated with the retrieved prior \(c_{\mathrm{prior}}\).

The retrieved prior may either \emph{match} or \emph{mismatch} the current graph-question pair. A valid match requires both operation compatibility and representation coverage. Let \((u_{s^\star},\tau_{s^\star})\) denote the decoded condition associated with \(h_{s^\star}\). We define the matching indicator as:
\begin{equation}
m=
\mathbb{I}\!\left[
      \tau=\tau_{s^\star}
  \;\land\;
  \mathrm{Cov}_{\mathrm{rep}}(u,u_{s^\star})=1
\right],
\end{equation}
where \(\mathrm{Cov}_{\mathrm{rep}}(u,u_{s^\star})\) checks whether the representation capability encoded by the retrieved prior covers the topology and state requirements implied by the current representation cues \(u\).

\textit{Matched}. When \(m=1\), the retrieved prior is regarded as compatible with the current graph-question condition. In this case, we directly reuse \(c_{\mathrm{prior}}\) as the final code instance for execution.

\textit{Mismatched}. When \(m=0\), the retrieved prior does not cover the current condition. This occurs when the operation cue is inconsistent (e.g., the current query requires shortest-path execution while the retrieved prior corresponds to max-flow computation), or when the representation capability is insufficient (e.g., the current graph is weighted while the retrieved prior only supports unweighted graph representation). In this case, we use the retrieved prior only as contextual guidance and let the VLM write a new code instance conditioned on the current input, the current signature, and the mismatched prior:
\begin{equation}
c=\mathrm{VLM}_{\theta}(x,q,h,c_{\mathrm{prior}}).
\end{equation}
This inference process is designed to serve as a reusable prior when conditions are compatible, while still enabling the model to generate new code when prior reuse is not sufficient.

\subsection{Layered Objective for Representation and Operation (LORO)}

We model visual graph reasoning as generating a paired compiler code instance for representation recovery and downstream operation, and train the code policy with reinforcement learning guided by a layered reward, denoted as LORO. Instead of rewarding final answers only, LORO applies reward in layers aligned with the task structure: the code must first be executable, then produce a valid compiled graph object, and only then receive semantic credit for representation and operation quality. This design reduces ambiguous credit assignment between representation quality and operation quality, and suppresses shortcut solutions built on incomplete compiled representations.

Conditioned on the current graph-question pair and the observer signature, we denote by \(\pi_\theta(\cdot\mid x,q,h)\) the code-generation policy induced by the VLM parameterized by \(\theta\), from which we sample \(G\) candidate code instances:
\begin{equation}
c^{(1)},\ldots,c^{(G)}
\sim
\pi_\theta\!\left(\cdot \mid x,q,h\right).
\end{equation}

For a candidate paired code instance \(c^{(g)}\), the system first executes \(c^{\mathrm{rep},(g)}\) to produce a compiled representation \(z^{(g)}\), and then executes \(c^{\mathrm{op},(g)}\) to produce a final answer \(\hat y^{(g)}\). We use \(E^{(g)}\in\{0,1\}\) to indicate whether the code is executable, and \(V^{(g)}\in\{0,1\}\) to indicate whether the compiled representation can be instantiated as a valid graph object under the target schema, regardless of whether the recovered graph is semantically correct. For valid code, we further define:
\begin{equation}
C^{(g)}=
\frac{1}{|\mathcal M(x,q)|}
\sum_{m\in \mathcal M(x,q)}
\mathrm{F1}_{m}^{(g)},
\qquad
O^{(g)}=\mathcal J_q(\hat y^{(g)},y^\star),
\end{equation}

\noindent where \(\mathcal M(x,q)\) denotes the set of topology and state attributes that must be recovered for the current graph-question pair, such as node set, edge set, direction, and edge weight; \(\mathrm{F1}_{m}^{(g)}\) is the F1 score on attribute \(m\); and \(\mathcal J_q(\cdot,\cdot)\in[0,1]\) is a task-specific operation score normalized to \([0,1]\), with \(y^\star\) denoting the reference answer. In particular, \(\mathcal J_q\) is exact match for binary questions, and a normalized score for scalar outputs such as shortest path length.

LORO defines the reward for candidate \(g\) as:
\begin{equation}
r^{(g)}=E^{(g)}V^{(g)}\!\left(C^{(g)} + C^{(g)} O^{(g)}\right) - 1.
\end{equation}

The \(-1\) offset assigns a common failure penalty \(r^{(g)}=-1\) to unusable code, keeps rewards in \([-1,1]\), and makes positive reward possible only when representation recovery and downstream operation jointly reach sufficient quality. The factor \(E^{(g)}V^{(g)}\) gates semantic credit to executable code that produces a valid compiled graph object, preventing malformed compilations with nontrivial attribute-level F1 from receiving reward. Within this valid branch, \(C^{(g)}\) rewards faithful representation recovery, while \(C^{(g)}O^{(g)}\) rewards downstream operation only through representation quality. This coupling suppresses shortcut codes that solve the query from a small local fragment and instead drives optimization toward faithful graph recovery before downstream task success.

For the \(G\) candidates sampled from the same input, we compute group-relative advantages directly from their rewards:
\begin{equation}
\hat A^{(g)}=
\frac{
r^{(g)}-\mathrm{mean}\!\left(r^{(1:G)}\right)
}{
\mathrm{std}\!\left(r^{(1:G)}\right)+\epsilon
}.
\end{equation}

We then update the code policy with the following objective, where \(\epsilon\) is a small constant for numerical stability, \(\beta\) is the KL regularization coefficient, and \(\pi_{\mathrm{ref}}\) is the reference policy:
\begin{equation}
\mathcal L(\theta)=
-\frac{1}{G}\sum_{g=1}^{G}
\hat A^{(g)}
\log \pi_\theta\!\left(c^{(g)}\mid x,q,h\right)
+\beta\,\mathrm{KL}\!\left(\pi_\theta\|\pi_{\mathrm{ref}}\right).
\end{equation}

\subsection{Code Archive and Inference}

During training, VGCompiler pairs a frozen observer with a trainable code policy. In our main experiments, both are instantiated from Qwen3-VL-8B~\cite{qwen3vl8b}. For each graph-question pair, the observer produces a condition signature, and the code policy samples a group of \(G\) candidate code instances. LORO evaluates these candidates by checking executability and compiled graph validity, then assigning representation and operation rewards within the valid branch before updating the policy with the group-relative objective. In parallel, VGCompiler accumulates stable code under similar graph-question conditions into a code archive, without using the archive to assist the current training sample:
\begin{equation}
\mathcal C^\star=\{(h_s,\hat c_s,\bar R_s)\}_{s=1}^{S},
\end{equation}
where \(S\) is the archive size, \(h_s\) is the condition signature, \(\hat c_s\) is the retained code, and \(\bar R_s\) is its average validation reward. In practice, we group candidate code by condition signature, retain only executable candidates with valid compiled graphs, and keep the highest-reward code as the archive entry.

During inference, both the observer and the code policy are frozen, and the fixed code archive is queried for reuse. Given a new graph-question pair \((x,q)\), VGCompiler computes its condition signature, retrieves the nearest code prior, checks representation coverage and operation compatibility, and either directly executes the matched prior or generates a new code instance conditioned on the current input and the mismatched prior.

\begin{table*}[t]
\centering
\setlength{\abovecaptionskip}{2pt}
\setlength{\belowcaptionskip}{0pt} 
\caption{Main results on three public visual-graph benchmarks. Vis.-Only denotes visual graph input only, Hybrid VL denotes the setting that combines visual input with adjacency-list input. We report mean \(\pm\) standard deviation over five runs.}
\label{tab:main_results_public_benchmarks_recomputed_from_5runs}
\setlength{\tabcolsep}{3pt}
\renewcommand{\arraystretch}{0.9}
\small
\begin{tabular}{lccccccccc}
\toprule
\multirow{2}{*}{Method} & \multicolumn{2}{c}{GVLQA-Base} & \multicolumn{2}{c}{GVLQA-AUG} & \multicolumn{2}{c}{VisionGraph} & \multicolumn{2}{c}{VGCURE} & \multirow{2}{*}{Average} \\
\cmidrule(lr){2-3}\cmidrule(lr){4-5}\cmidrule(lr){6-7}\cmidrule(lr){8-9}
& Vis.-Only & Hybrid VL & Vis.-Only & Hybrid VL & Node Rec. & Edge Rec. & Graph Und. & Graph Reas. & \\
\midrule

\rowcolor{black!10}
\multicolumn{10}{l}{\textbf{Closed-source VLMs}} \\
GPT-5-mini & $49.40 \pm 0.98$ & $45.20 \pm 1.00$ & $46.40 \pm 4.81$ & $47.60 \pm 2.80$ & $85.70 \pm 1.92$ & $1.80 \pm 0.07$ & $75.00 \pm 3.45$ & $60.11 \pm 2.62$ & $50.29 \pm 2.11$ \\
GPT-5.2 & $53.60 \pm 5.51$ & $45.20 \pm 4.79$ & $59.51 \pm 7.75$ & $51.20 \pm 5.49$ & $91.70 \pm 1.03$ & $14.30 \pm 1.68$ & $85.72 \pm 1.75$ & $68.51 \pm 3.80$ & $57.55 \pm 3.96$ \\
Gemini-3.1-Flash & $54.80 \pm 8.95$ & $53.00 \pm 12.63$ & $47.60 \pm 8.56$ & $56.00 \pm 15.77$ & $91.10 \pm 0.06$ & $6.00 \pm 0.92$ & $83.92 \pm 0.11$ & $\underline{77.40 \pm 1.07}$ & $57.46 \pm 6.01$ \\
Gemini-3.1-pro & $\underline{59.11 \pm 5.85}$ & $52.00 \pm 6.78$ & $\underline{62.51 \pm 8.54}$ & $57.40 \pm 9.46$ & $\bm{96.88 \pm 0.73}$ & $18.00 \pm 1.41$ & $\underline{90.68 \pm 1.25}$ & $75.51 \pm 2.98$ & $62.83 \pm 4.62$ \\
Claude-Haiku-4.5 & $20.20 \pm 1.00$ & $48.80 \pm 0.96$ & $24.40 \pm 1.07$ & $44.00 \pm 1.01$ & $69.60 \pm 7.63$ & $0.03 \pm 0.06$ & $70.80 \pm 1.03$ & $61.91 \pm 3.71$ & $41.18 \pm 2.15$ \\
Claude-Sonnet-4.5 & $49.86 \pm 4.16$ & $47.20 \pm 3.46$ & $55.51 \pm 5.83$ & $51.51 \pm 3.95$ & $89.72 \pm 3.03$ & $13.53 \pm 1.15$ & $84.84 \pm 1.65$ & $68.94 \pm 3.86$ & $56.45 \pm 3.40$ \\
\midrule

\rowcolor{black!10}
\multicolumn{10}{l}{\textbf{Open-source VLMs}} \\
LLaVA-1.5-7B & $9.51 \pm 2.10$ & $11.31 \pm 4.16$ & $11.91 \pm 1.56$ & $11.91 \pm 2.10$ & $3.58 \pm 1.37$ & $0.10 \pm 0.10$ & $16.08 \pm 3.61$ & $14.29 \pm 1.36$ & $9.33 \pm 1.94$ \\
LLaVA-1.5-13B & $10.80 \pm 2.01$ & $13.37 \pm 3.82$ & $13.20 \pm 1.74$ & $13.66 \pm 2.15$ & $4.20 \pm 1.44$ & $0.07 \pm 0.11$ & $18.24 \pm 3.04$ & $16.51 \pm 1.52$ & $10.68 \pm 1.89$ \\
InternVL3.5-8B & $32.14 \pm 1.47$ & $20.83 \pm 3.91$ & $29.17 \pm 5.81$ & $26.20 \pm 4.20$ & $63.70 \pm 2.33$ & $0.12 \pm 0.12$ & $61.32 \pm 0.81$ & $55.94 \pm 2.28$ & $35.13 \pm 2.63$ \\
Qwen3-VL-8B & $32.74 \pm 2.40$ & $23.80 \pm 0.82$ & $34.51 \pm 0.87$ & $23.20 \pm 1.41$ & $73.20 \pm 0.07$ & $0.70 \pm 0.78$ & $75.00 \pm 0.14$ & $58.34 \pm 2.40$ & $38.83 \pm 1.11$ \\
\midrule

\rowcolor{black!10}
\multicolumn{10}{l}{\textbf{Direct-visual methods}} \\
GITA-7B & $17.86 \pm 0.68$ & $21.37 \pm 0.73$ & $20.83 \pm 0.26$ & $22.11 \pm 0.27$ & $16.57 \pm 0.65$ & $0.12 \pm 0.15$ & $22.00 \pm 0.32$ & $19.29 \pm 0.29$ & $17.03 \pm 0.42$ \\
GITA-13B & $16.86 \pm 0.71$ & $24.77 \pm 0.67$ & $20.26 \pm 0.27$ & $25.60 \pm 0.31$ & $16.80 \pm 0.62$ & $0.15 \pm 0.16$ & $28.00 \pm 0.37$ & $24.00 \pm 0.32$ & $18.86 \pm 0.42$ \\
MCDGraph-8B & $48.80 \pm 3.89$ & $39.29 \pm 1.83$ & $42.26 \pm 1.02$ & $47.63 \pm 8.61$ & $85.12 \pm 3.56$ & $5.35 \pm 2.87$ & $72.04 \pm 1.08$ & $63.11 \pm 3.89$ & $49.49 \pm 3.42$ \\
\midrule

\rowcolor{black!10}
\multicolumn{10}{l}{\textbf{Code-based methods}} \\
DPR + GPT-5.2 & $51.20 \pm 0.95$ & $\underline{91.06 \pm 0.41}$ & $47.63 \pm 4.86$ & $\underline{98.43 \pm 2.56}$ & $86.30 \pm 3.48$ & $\underline{37.75 \pm 1.92}$ & $75.60 \pm 0.95$ & $60.11 \pm 2.55$ & $\underline{68.03 \pm 2.24}$ \\
Qwen3-VL-8B + Code & $31.20 \pm 1.03$ & $82.06 \pm 0.44$ & $36.63 \pm 5.26$ & $87.80 \pm 5.99$ & $79.10 \pm 5.04$ & $23.00 \pm 1.96$ & $60.32 \pm 0.98$ & $58.46 \pm 2.67$ & $56.99 \pm 3.00$ \\
Qwen-VGCompiler & $\bm{91.09 \pm 1.04}$ & $\bm{97.20 \pm 1.06}$ & $\bm{88.23 \pm 0.33}$ & $\bm{98.91 \pm 1.00}$ & $\underline{96.30 \pm 0.78}$ & $\bm{77.75 \pm 3.74}$ & $\bm{95.64 \pm 0.33}$ & $\bm{91.11 \pm 1.09}$ & $\bm{91.72 \pm 1.23}$ \\
\bottomrule
\end{tabular}
\end{table*}

\section{Experiments}

\subsection{Experimental Setup}

\textbf{Benchmarks and data splits.} We train and evaluate VGCompiler on three public visual graph reasoning benchmarks: GVLQA from GITA~\cite{gita2024}, VisionGraph~\cite{visiongraph2024}, and VGCURE~\cite{vgcure2025}. GVLQA is reported on \textit{GVLQA-Base}, the clean benchmark with seven graph-theoretic tasks, and \textit{GVLQA-AUG}, which keeps the same tasks but introduces rendering perturbations. Both are evaluated under \textit{Vis.-Only}, which uses only the rendered graph, and \textit{Hybrid VL}, which additionally provides the adjacency-list input. For VisionGraph, we report \textit{Node Recognition} and \textit{Edge Recognition}, which test node counting and full edge-set recovery. For VGCURE, we report \textit{Graph Understanding}, which covers basic structural queries, and \textit{Graph Reasoning}, which covers relation-based and path-based reasoning over entity-relation graphs. For training and evaluation, we follow the default settings of the three benchmarks.

\textbf{Baselines.}
We compare VGCompiler with four groups of baselines: (1) Closed-source VLMs, including GPT-5-mini~\cite{gpt5}, GPT-5.2~\cite{gpt5}, Gemini-3.1-Flash~\cite{gemini31}, Gemini-3.1-pro~\cite{gemini31}, Claude-Haiku-4.5~\cite{claude45}, and Claude-Sonnet-4.5~\cite{claude45}; (2) Open-source VLMs, including LLaVA-1.5-7B~\cite{llava15}, LLaVA-1.5-13B~\cite{llava15}, InternVL3.5-8B~\cite{internvl35}, and Qwen3-VL-8B~\cite{qwen3vl8b}; (3) Direct-visual methods, including GITA~\cite{gita2024}, which is trained on top of LLaVA-1.5-7B~\cite{llava15} and LLaVA-1.5-13B~\cite{llava15}, and MCDGraph~\cite{vgcure2025}, which we reproduce with Qwen3-VL-8B~\cite{qwen3vl8b}; and (4) Code-based methods, including Description Programming Reasoning (DPR)~\cite{visiongraph2024}, instantiated with GPT-5.2~\cite{gpt5}, Qwen3-VL-8B + Code based on Qwen3-VL-8B~\cite{qwen3vl8b}, a prompt-only control that uses our prompting scheme without training and without code-archive retrieval, and our Qwen-VGCompiler.

\textbf{Evaluation protocol.}
We use exact-match Accuracy as the primary metric, reported as percentages and shown as mean \(\pm\) standard deviation over five runs.

\textbf{Implementation details.}
Both the trainable code policy and the frozen observer are initialized from Qwen3-VL-8B~\cite{qwen3vl8b}. During training, we sample \(G=8\) candidate programs per input and maintain a key-value code archive indexed by condition signatures, which is queried only at inference by Hamming-distance retrieval. At inference, the observer uses temperature 0 and the code policy uses temperature 0.7. Both training and inference are conducted on \(8\times\) NVIDIA A800 80GB GPUs. We set \(\beta=0.01\), with \(\pi_{\mathrm{ref}}\) taken as a frozen copy of the initial code policy before LORO training.

Our experiment is designed to answer the following three research questions:
\begin{itemize}[leftmargin=*]
\item \textbf{RQ1 (Main Results)}: How effective and efficient is VGCompiler on public visual graph reasoning benchmarks?
\item \textbf{RQ2 (Ablation and Mechanism Analysis)}: Why does VGCompiler work, how do its key designs contribute to the gains?
\item \textbf{RQ3 (Generalization and Case Analysis)}: How does VGCompiler generalize beyond synthetic benchmarks and reuse code?

\end{itemize}

\begin{figure*}[t]
\centering
\setlength{\abovecaptionskip}{2pt}
\setlength{\belowcaptionskip}{0pt} 
\includegraphics[width=0.95\textwidth]{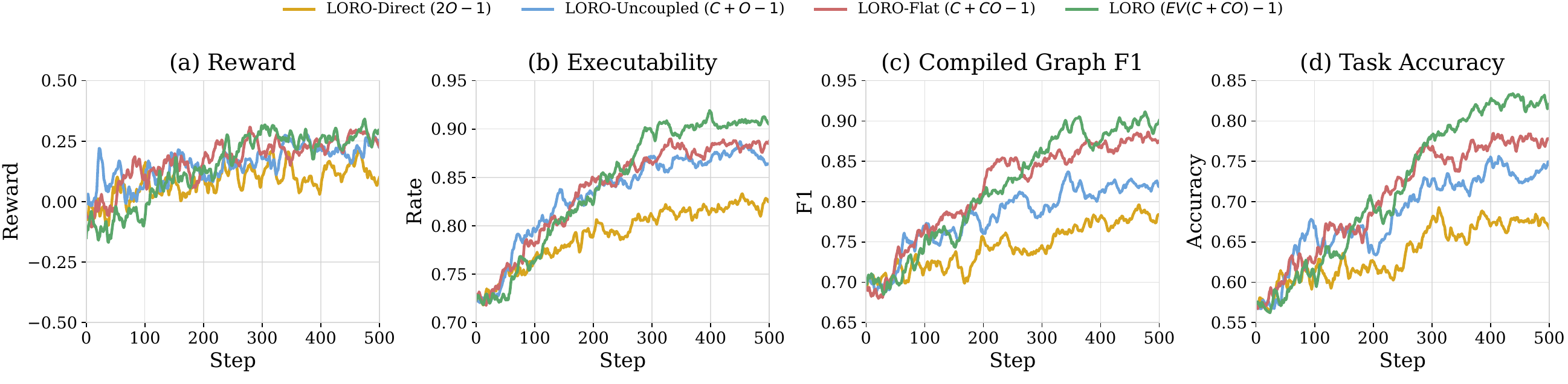}
\caption{Training dynamics of four optimization objectives. We report smoothed reward, executable rate, compiled graph F1, and final task accuracy during training. LORO-Direct uses result-only reward, LORO-Uncoupled removes representation-operation coupling, LORO-Flat removes executability-and-validity gating, and LORO uses the full layered reward.}
\label{fig:loco_training_curve}
\end{figure*}

\subsection{Main Results (RQ1)}

Table~\ref{tab:main_results_public_benchmarks_recomputed_from_5runs} reports the main results on three public benchmarks. Overall, VGCompiler achieves the best average accuracy, surpassing the strongest closed-source VLM baseline by 28.89\% and the closed-source code-based baseline DPR by 23.69\%, despite using an 8B open-source backbone. We draw the following observations:

\begin{itemize}[leftmargin=*,noitemsep]
\item VGCompiler outperforms all baselines across the benchmarks. The gain is largest when representation recovery is the main bottleneck, especially on VisionGraph Edge Recognition, which requires exact recovery of the full graph topology. This shows that a major source of VGCompiler's advantage lies in improving representation faithfulness from visual input.

\item The gain does not come from code generation alone. Although Qwen3-VL-8B + Code outperforms direct visual methods in several settings, it remains unstable and can even fall behind direct Qwen3-VL-8B on some tasks. This indicates that generated code is useful only when the compiled graph is sufficiently faithful.

\item The advantage of VGCompiler is not limited to representation recovery. Its lead remains clear on reasoning settings that require state-matched downstream operation, rather than only graph understanding, showing that faithful compilation must be paired with valid operation on the recovered graph state.

\item Compared with prior code-based pipeline DPR, VGCompiler is stronger not only because it compiles graphs more faithfully, but also because it more reliably couples representation recovery with downstream graph operation. This is especially important for tasks whose answers are sensitive to edge completeness, direction, weights, and the validity of the executed operation.
\end{itemize}

\begin{table}[t]
\centering
\setlength{\abovecaptionskip}{2pt}
\setlength{\belowcaptionskip}{0pt} 
\caption{Inference efficiency on the public benchmarks. For code-based methods, Time includes both model generation and code execution; for closed-source VLMs, the Input/Output tokens are taken from API usage statistics.}
\label{tab:efficiency}
\small
\renewcommand{\arraystretch}{0.95}
\begin{tabular*}{\columnwidth}{@{\extracolsep{\fill}}lcccc}
\toprule
Method & Accuracy & In Tokens & Out Tokens & Time \\
\midrule
GPT-5-mini & 50.29 & 1.89k & 0.05k & 4.8s \\
GPT-5.2 & 57.55 & 1.91k & 0.13k & 7.3s \\
Gemini-3.1-pro & 62.83 & 1.17k & 1.02k & 13.1s \\
DPR + GPT-5.2 & 68.03 & 5.26k & 3.34k & 18.9s \\
Qwen3-VL-8B + Code & 56.99 & 3.01k & 1.91k & 12.8s \\
Qwen-VGCompiler & \textbf{91.72} & 3.98k & 1.10k & 7.7s \\
\bottomrule
\end{tabular*}
\end{table}

\begin{table}[t]
\centering
\setlength{\abovecaptionskip}{2pt}
\setlength{\belowcaptionskip}{0pt} 
\caption{Ablation on target language. We report task accuracy, executable rate, compiled graph validity, and compiled graph F1 denote as Accuracy, \(E\), \(V\), and F1, respectively.}
\label{tab:design_ablation}
\setlength{\tabcolsep}{2.5pt}
\renewcommand{\arraystretch}{0.95}
\small
\begin{tabular*}{\columnwidth}{@{\extracolsep{\fill}}@{}lcccc@{}}
\toprule
Target language & Accuracy & \(E\) & \(V\) & F1 \\
\midrule
Adjacency matrix & 82.7 & 89.7 & 85.6 & 76.5 \\
Edge list & 84.8 & 94.0 & 91.4 & 80.3 \\
Attr. adjacency list & \textbf{91.7} & \textbf{96.8} & \textbf{92.8} & \textbf{94.6} \\
\bottomrule
\end{tabular*}
\end{table}

Table~\ref{tab:efficiency} shows that VGCompiler also has a clear inference efficiency advantage. Compared with Gemini-3.1-pro, it achieves substantially lower latency, suggesting that code execution can be more efficient than relying on heavy closed-source internal reasoning. Compared with DPR, VGCompiler is faster because DPR depends on a closed-source model to recover graph structure through staged prompting, whereas VGCompiler uses a lightweight VLM observer to extract cues and delegates the downstream graph operation to code execution. Compared with Qwen3-VL-8B + Code, VGCompiler slightly increases input tokens due to archive retrieval and prior code prompting, but reduces output tokens and latency, leading to a better overall efficiency-accuracy trade-off.

\subsection{Ablation and Mechanism Analysis (RQ2)}

\begin{table}[t]
\centering
\setlength{\abovecaptionskip}{2pt}
\setlength{\belowcaptionskip}{0pt} 
\caption{Factorizing the effect of archive priors and LORO on the public benchmarks.}
\label{tab:factorization}
\small
\renewcommand{\arraystretch}{0.95}
\begin{tabular*}{\columnwidth}{@{\extracolsep{\fill}}lcccc}
\toprule
Variant & Accuracy & \(E\) & \(V\) & F1 \\
\midrule
Qwen3-VL-8B + Code & 57.0 & 72.3 & 74.1 & 69.8 \\
VGCompiler w/o Archive & 80.1 & 91.0 & 88.5 & 89.1 \\
VGCompiler w/o LORO & 71.6 & 94.1 & 85.4 & 87.4 \\
VGCompiler & \textbf{91.7} & \textbf{96.8} & \textbf{92.8} & \textbf{94.6} \\
\bottomrule
\end{tabular*}
\end{table}

\begin{figure}[t]
\centering
\setlength{\abovecaptionskip}{2pt}
\setlength{\belowcaptionskip}{0pt} 
\includegraphics[width=\columnwidth]{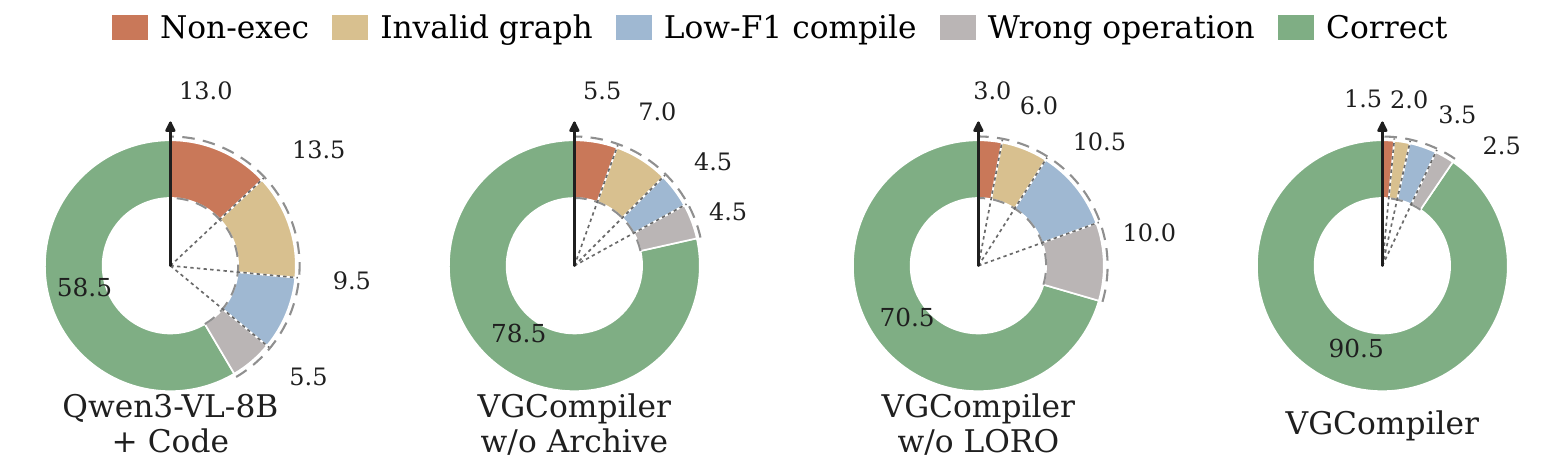}
\caption{Error analysis on 200 sampled cases. We classify errors into four categories: non executable code, invalid compiled graph, low fidelity compilation, and wrong operation.}
\label{fig:error_breakdown}
\end{figure}

Table~\ref{tab:design_ablation} shows that the advantage of the attributed adjacency list does not mainly come from representational completeness. All three target languages can encode the required graph semantics, and adjacency matrices and edge lists already achieve relatively high compiled graph validity under the current definition of \(V\). The key difference lies in how the target language shapes compilation and operation. Adjacency matrices encourage one-shot global recovery and are therefore more sensitive to visual ambiguity and style variation. Edge lists recover the graph edge by edge, making them more vulnerable to missing local relations that break the semantics required by reasoning. By contrast, the attributed adjacency list preserves local neighborhood structure while directly exposing direction, weight, and state in executable form, leading to the strongest compiled graph F1 and final task accuracy.

Figure~\ref{fig:loco_training_curve} compares four optimization objectives to verify the effect of LORO. LORO-Direct uses a result-only reward \(2O-1\), LORO-Uncoupled uses \(C+O-1\), LORO-Flat uses \(C+CO-1\), and LORO uses \(EV(C+CO)-1\). The curves show that LORO-Direct improves reward and executability early on, but soon saturates in compiled graph F1 and final task accuracy because it is too weak to sustain faithful graph recovery. LORO-Uncoupled already improves both metrics, showing the importance of directly rewarding representation quality, but its operation reward remains independent of representation fidelity and can still favor codes with valid compiled graphs but low fidelity. LORO-Flat further improves compiled graph F1 and final task accuracy, indicating that coupling operation reward to representation fidelity is the key step for suppressing such shortcut behavior. Full LORO then further restricts semantic credit to executable code with valid compiled graph objects, preventing malformed but answer-sufficient programs from receiving reward. These results show that LORO improves visual graph reasoning by combining faithful compilation with validity-aware and operation-aware optimization.

\begin{table}[t]
\centering
\setlength{\abovecaptionskip}{2pt}
\setlength{\belowcaptionskip}{0pt} 
\caption{Archive concentration for path-style queries. Top-1 and Top-2 report dominant-signature coverage, and main split cues summarize the factors separating signature groups.}
\label{tab:strategy_brief}
\footnotesize
\setlength{\tabcolsep}{1.2pt}
\begin{tabular*}{\columnwidth}{@{\extracolsep{\fill}}lp{0.065\columnwidth}p{0.085\columnwidth}p{0.085\columnwidth}p{0.39\columnwidth}@{}}
\toprule
Family & \#Sig. & Top-1 & Top-2 & Main split cues \\
\midrule
GVLQA-Base & 4 & 47\% & 79\% & Text support; Edge.density \\
GVLQA-AUG & 11 & 19\% & 37\% & Node.style; Edge.thickness \\
VisionGraph & 7 & 27\% & 50\% & Node.density; Weight.placement \\
VGCURE & 6 & 31\% & 59\% & Edge.label; Direction \\
\bottomrule
\end{tabular*}
\vspace{-1em}
\end{table}

Table~\ref{tab:factorization} and Figure~\ref{fig:error_breakdown} jointly clarify how archive reuse and LORO contribute to VGCompiler. Relative to Qwen3-VL-8B + Code, introducing LORO already brings substantial gains in compiled graph validity, compiled graph F1, and final task accuracy. Removing the archive from VGCompiler mainly lowers executability, while removing LORO more directly influences compiled graph validity and final task accuracy. Figure~\ref{fig:error_breakdown} shows the same pattern at the error level: removing the archive mainly increases non executable and invalid compiled graph errors, whereas removing LORO mainly increases low fidelity compilation and wrong operation errors. Together, these results suggest that archive reuse mainly stabilizes executable code under matched conditions, while LORO provides the stronger semantic optimization over faithful compilation and downstream operation.

\subsection{Generalization and Case Analysis (RQ3)}

Table~\ref{tab:strategy_brief} shows that the code archive does not collapse each benchmark family into a single universal program. Instead, each family is covered by a small set of dominant signatures. GVLQA-Base is the most concentrated, mainly split by text support and edge density, whereas GVLQA-AUG is the most diverse because style perturbations expand the signature space through node appearance and edge thickness. VisionGraph remains relatively diverse because dense weighted renderings require distinct representation routes, while VGCURE is mainly separated by relation labels and direction cues. These results suggest archive retrieval is organized around a small number of reusable but non-universal graph-reasoning programs.

We further evaluate whether VGCompiler generalizes beyond synthetic benchmarks to real-world graph diagrams. To this end, we build a real-world benchmark with three domains: metro routing, logistics delivery, and network fault assessment, and evaluate VGCompiler directly without training. Each task requires both faithful graph representation recovery and a domain-grounded downstream operation. Metro reasoning requires least-transfer routing, logistics reasoning requires capacity-constrained warehouse assignment, and network reasoning requires single-point network fault assessment.

Table~\ref{tab:real_world_reasoning} shows that Qwen-VGCompiler achieves the best performance in both final task accuracy and compiled graph F1 across these heterogeneous real-world settings, outperforming all baselines. Compared with the strongest baseline DPR, the gains remain consistent across all three domains, despite their different visual conventions and operational objectives. This suggests that the advantage of VGCompiler is not limited to benchmark-specific patterns, but comes from a more general ability to recover structure-preserving graph representations from visually diverse diagrams and support valid downstream operations reliably. Logistics remains the hardest setting in Accuracy for all methods, which is consistent with the fact that warehouse assignment still requires reliable state recovery together with capacity-aware customer allocation. Meanwhile, network fault assessment provides a complementary challenge, as heterogeneous device types and semantically richer links make the mapping from visual evidence to graph structure and downstream operation more explicit. Figure~\ref{fig:path_signature_example} further provides a concrete real-world case from network fault assessment, showing how VGCompiler compiles device connectivity from the diagram and then executes final analysis.

\begin{table}[t]
\centering
\setlength{\abovecaptionskip}{2pt}
\setlength{\belowcaptionskip}{0pt} 
\caption{Results on the real-world generalization benchmark. We report final task accuracy and compiled graph F1.}
\label{tab:real_world_reasoning}
\setlength{\tabcolsep}{2.5pt}
\footnotesize
\resizebox{\columnwidth}{!}{
\begin{tabular}{lcccccc}
\toprule
\multirow{2}{*}{Method} & \multicolumn{2}{c}{Metro} & \multicolumn{2}{c}{Logistics} & \multicolumn{2}{c}{Network} \\
\cmidrule(lr){2-3}\cmidrule(lr){4-5}\cmidrule(lr){6-7}
&Accuracy  & Graph F1 & Accuracy & Graph F1 & Accuracy& Graph F1 \\
\midrule
GPT-5.2 & $45.7 \pm 2.1$ & $63.8 \pm 1.6$ & $30.5 \pm 1.8$ & $46.5 \pm 1.7$ & $43.2 \pm 2.3$ & $59.7 \pm 2.0$ \\
Gemini-3.1-pro & $48.2 \pm 2.4$ & $65.1 \pm 1.8$ & $33.5 \pm 2.0$ & $49.1 \pm 1.8$ & $46.3 \pm 1.6$ & $62.4 \pm 1.6$ \\
InternVL3.5-8B & $19.2 \pm 1.6$ & $34.8 \pm 1.7$ & $11.8 \pm 1.2$ & $27.8 \pm 1.1$ & $23.7 \pm 1.9$ & $39.5 \pm 1.5$ \\
Qwen3-VL-8B & $23.8 \pm 1.2$ & $39.7 \pm 1.1$ & $15.0 \pm 1.3$ & $31.7 \pm 1.2$ & $27.3 \pm 1.2$ & $42.7 \pm 1.3$ \\
GITA-7B & $11.3 \pm 0.8$ & $25.9 \pm 0.9$ & $8.2 \pm 0.6$ & $20.4 \pm 0.7$ & $14.8 \pm 1.0$ & $29.2 \pm 1.0$ \\
DPR + GPT-5.2 & $\underline{61.7 \pm 1.4}$ & $\underline{77.3 \pm 1.0}$ & $\underline{45.5 \pm 1.8}$ & $\underline{60.4 \pm 1.4}$ & $\underline{58.0 \pm 1.7}$ & $\underline{72.6 \pm 1.2}$ \\
Qwen-VGCompiler & $\bm{90.3 \pm 0.8}$ & $\bm{91.8 \pm 0.8}$ & $\bm{63.5 \pm 1.5}$ & $\bm{81.2 \pm 1.2}$ & $\bm{70.0 \pm 1.3}$ & $\bm{82.2 \pm 1.1}$ \\
\bottomrule
\end{tabular}
}
\end{table}

\begin{figure}[t]
\centering
\setlength{\abovecaptionskip}{2pt}
\setlength{\belowcaptionskip}{0pt} 
\includegraphics[width=\columnwidth]{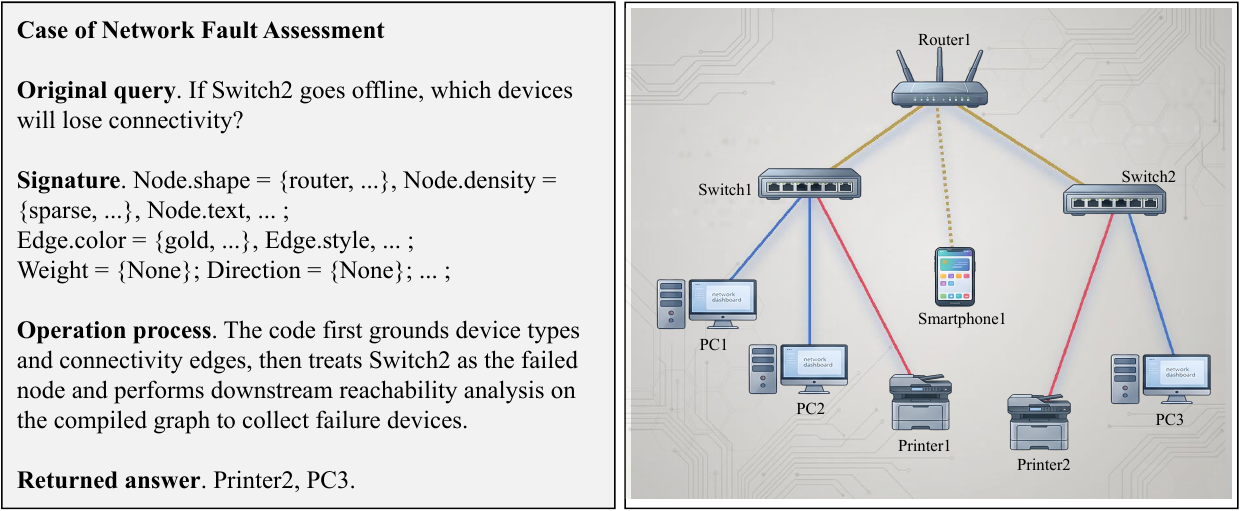}
\caption{A real-world case from network fault assessment. The left panel summarizes the query, signature, operation process, and returned answer, while the right panel shows the corresponding network diagram.}
\label{fig:path_signature_example}
\end{figure}

\section{Conclusions}
We present VGCompiler, a compilation-centric paradigm for visual graph reasoning under a knowledge compilation view. VGCompiler organizes reasoning through graph representation recovery and valid downstream operation. It further combines condition-aware code retrieval with reinforcement learning guided by a layered reward over executability, compiled graph validity, representation quality, and operation quality. Across three public benchmarks and real-world scenarios, VGCompiler consistently outperforms prior methods while remaining efficient. The results suggest that treating visual graph reasoning as a knowledge compilation process with reusable executable code provides an effective foundation for robust reasoning over heterogeneous visual graphs.

\balance
\bibliographystyle{ACM-Reference-Format}
\bibliography{reference}

\newpage

\end{document}